\documentclass[%
 reprint,
 superscriptaddress,
 amsmath,
 amssymb,
 longbibliography
]{revtex4-1}

\usepackage{graphicx}
\usepackage{dcolumn}
\usepackage{bm}
\usepackage{tikz}
\usepackage{qcircuit}
\usepackage{braket}
\usepackage{appendix}
\usepackage{xcolor}
\usepackage[caption=false]{subfig}

\begin{document}

\preprint{APS/123-QED}

\title{Cost function embedding and dataset encoding for \\ machine learning with parameterized quantum circuits }

\author{Shuxiang Cao}
\email{shuxiang.cao@physics.ox.ac.uk}
\affiliation{Department of Physics, Clarendon Laboratory, University of Oxford, OX1 3PU, UK}
\affiliation{Rahko Limited, London N4 3JP, UK}
\author{Leonard Wossnig}
\affiliation{Department of Computer Science, University College London, London WC1E 6BT, UK}
\affiliation{Rahko Limited, London N4 3JP, UK}
\author{Brian Vlastakis}
\affiliation{Department of Physics, Clarendon Laboratory, University of Oxford, OX1 3PU, UK}
\affiliation{Oxford Quantum Circuits Limited, Oxford OX2 6HT, UK}
\author{Peter Leek}
\affiliation{Department of Physics, Clarendon Laboratory, University of Oxford, OX1 3PU, UK}
\affiliation{Oxford Quantum Circuits Limited, Oxford OX2 6HT, UK}
\author{Edward Grant}
\affiliation{Department of Computer Science, University College London, London WC1E 6BT, UK}
\affiliation{Rahko Limited, London N4 3JP, UK}

\begin{abstract}
Machine learning is seen as a promising application of quantum computation. For near-term noisy intermediate-scale quantum (NISQ) devices, parametrized quantum circuits (PQCs) have been proposed as machine learning models due to their robustness and ease of implementation. However, the cost function is normally calculated classically from repeated measurement outcomes, such that it is no longer encoded in a quantum state. This prevents the value from being directly manipulated by a quantum computer. To solve this problem, we give a routine to embed the cost function for machine learning into a quantum circuit, which accepts a training dataset encoded in superposition or an easily preparable mixed state. We also demonstrate the ability to evaluate the gradient of the encoded cost function in a quantum state.

\end{abstract}

\maketitle

\section{Introduction}

Machine learning (ML) is one of the most successful research areas of the past decade and has a wide range of applications~\cite{LeCun2015DeepLearning, Guo2018ANetworks, Chiu2018State-of-the-ArtModels}.
Meanwhile, quantum computing is the area of research that aims to design more efficient or generally more powerful methods using a new model of computation based on quantum mechanics ~\cite{Nielsen2000QuantumInformation}.
Different methods to implement ML techniques on quantum computers have been proposed~\cite{Wiebe2014QuantumLearning, Rebentrost2014QuantumClassification, Du2018BayesianCircuit, Ciliberto2017QuantumPerspective, Biamonte2017QuantumLearning,Perdomo-Ortiz2018OpportunitiesComputers,Ciliberto2017QuantumPerspective}.
While the first stream of quantum ML algorithms exploited fast linear algebra subroutines to obtain a quantum speedup~\cite{Harrow2009QuantumEquations, Wossnig2018QuantumMatrices}, recently, classical-quantum hybrid approaches to ML, so-called parametrized quantum circuits (PQCs) have experienced a surge of interest~\cite{Grant2018HierarchicalClassifiers,Farhi2018ClassificationProcessors, Mitarai2018QuantumLearning,Benedetti2019ACircuits}. The main reason for their popularity is their suitability for Noisy Intermediate-Scale Quantum (NISQ) devices, hence their appeal to become a candidate for first applications in quantum ML~\cite{Grant2018HierarchicalClassifiers,Huggins2018TowardsNetworks, Perdomo-Ortiz2018OpportunitiesComputers}.

One exciting aspect of PQC-based ML is the equivalence to tensor-network-based ML algorithms~\cite{Huggins2018TowardsNetworks,Liu2019MachineStructure}.
Promising results indeed indicate that certain types of these algorithms can be executed efficiently on a quantum computer but cannot be efficiently evaluated classically, which implies that they have stronger expressive power \cite{Du2018TheCircuits}, and a more flexible feature space \cite{Havlicek2019SupervisedSpaces}.
Based on these results, there is a hope that such quantum models will yield a practical advantage in ML.

Just as with classical deep learning, parameter training is the most time-consuming process of building a PQC model. Finding efficient ways to obtain optimal or near-optimal parameters is a major objective of current research. Several interesting methods have been proposed to optimize parameters. Firstly, there are gradient-based optimization methods \cite{Harrow2019Low-depthAlgorithms}; the partial derivative of the expectation value can be directly evaluated with a Hadamard Test~\cite{Dallaire-Demers2019Low-depthComputer, Romero2018StrategiesAnsatz}, or the \textit{shift-rule} can be used to do a Hadamard Test with indirect measurement \cite{Mitarai2018QuantumLearning, Schuld2019EvaluatingHardware}. Alternatively, a gradient-free optimization method called \textit{rotosolve} can be used \cite{Ostaszewski2019QuantumLearning,Nakanishi2019SequentialAlgorithms}.

The target to be optimized (in ML, the \textit{cost function}) is typically calculated from measured expectation values. Therefore, the value of the cost function is no longer encoded in the quantum state. This brings us some inconvenience. For example, the Hadamard Test and shift-rule methods can be used together with the chain rule to calculate the cost-function gradient \cite{Baydin2017AutomaticSurvey}, but this requires several function evaluations
\footnote{Suppose our cost function is given by $f(x)$ and the measurement expectation value is given by $g(x)$, the derivative is then given by $f(g(x))^\prime = f^\prime(g(x)) g^\prime(x)$. To calculate this derivative, both the intermediate gradient $g(x)^\prime$ and the expectation value $g(x)$ need to be evaluated}.
In addition, the gradient-free method \textit{rotosolve}, which was developed for the specific value landscape of Hermitian expectation values, cannot be applied to classical non-linear cost functions \cite{Ostaszewski2019QuantumLearning,Nakanishi2019SequentialAlgorithms}.  

Therefore, an open question is whether we can find a method which embeds the cost function into a quantum circuit. Embedding the cost function into the quantum circuit directly will not only allow PQC-based ML models to be trained with the optimization techniques we mentioned above, but also opens the possibility for performing further quantum operations after the cost function is evaluated, such as quantum enhanced measurement \cite{Giovannetti2004Quantum-EnhancedLimit} and reduction of measurement times by phase estimation \cite{Wang2019AcceleratedEigensolver}. In this paper, we present a method which achieves the cost function embedding for PQC-based ML models.

We propose two cost functions that can be encoded directly through a quantum circuit, i.e., the outcome of the quantum circuit corresponds to the evaluation of the cost function for a given input. We then propose a corresponding data encoding method which allows the calculation of averaged cost function values among the encoded training dataset. 

\section{Classification with PQC}

Classification belongs to the category of supervised learning.
In supervised learning we are given a data set $\mathcal{S}:=\{x_i ,y_i\}_{i=1}^m$ of $m$ points, where $x_i \in \mathcal{X}$ are \textit{i.i.d.} samples drawn from a fixed but unknown probability distribution and $y_i \in \mathcal{Y}$ are the corresponding labels, i.e., the ideal output for a given input $x_i$.
In a classification task, the labels are obtained from a finite set of possible outcomes, for example, a binary outcome, i.e., $\mathcal{Y} = \{0,1\}$.
The goal of the classification task is then to find a function, the so-called model $f$, such that $f(x_i)$ returns a result that matches the label $y_i$ for all $i \in [m]$ with the highest possible accuracy on a subset of $\mathcal{S}$, called the training set $\mathcal{S}_t$.
Additionally we require that the function $f$ also performs well on samples $x_j \notin \mathcal{S}_t$ which we haven't used in order to train, i.e., find the model $f$, the so called unseen data or test set $\mathcal{S}_p = \mathcal{S}\setminus \mathcal{S}_t$.
A learning algorithm, or training algorithm $\mathcal{A}$, then takes as input the data set $\mathcal{S}$, and possibly additional hyper-parameters, and returns a model $f=\mathcal{A}(\mathcal{S},\cdot)$ which minimises some specified measure of the discrepancy between the desired output and the model output.
The measure of this discrepancy is called the cost function $\ell:\mathcal{Y} \times \mathcal{Y} \rightarrow \mathbb{R}$, also called the objective of the learning problem, and is an integral part of the solution.

In the PQC setting the model $f(x_i)$ is given by a large unitary matrix $U \in \mathbb{C}^{2^n \times 2^n}, U^\dag U=\mathbb{I}$ acting on an $n$-qubit system, and the input $x_i$ is given by a state $\ket{\psi_i}$.
A prediction of the model is then given by the expectation value for a measurement of one qubit after $U$ is applied to some input state $\ket{\psi_i} \in \mathbb{C}^{2^n}$, see Fig.~\ref{fig:cost_cnot}. Here we denote the qubits that store $\ket{\psi_i}$ as \textit{data qubits}.
$U$ can be decomposed in a number of different ways into elementary logical elements, which are called quantum gates.
One such decomposition then results in a sequence of single-qubit rotation gates and CNOT gates, which can then be parameterised by the rotation angles $\theta$ of the single-qubit gates~\cite{Vatan2004OptimalGates,Divincenzo1995Two-bitComputation}. 
After obtaining the gate sequence we arbitrarily choose a subset of the qubits as the \textit{output qubits}. The state of the output qubits is given by density matrix $\rho_{output}$. Then we consider their measurement output as the prediction of the ansatz.
For binary classification problems we choose only one of the qubits as the output and ignore the information stored in all others.
Once we have the output of the measurement of $\ket{\psi_i}$, i.e., the prediction of our model, the cost function can be evaluated by measurement and applying some post-processing in order to evaluate the discrepancy $\ell(f(\ket{\psi_i}), y_i)$.
The training task then consists of adjusting the parameters $\theta$ of the model $U(\theta)$ in order to match the expectation value of a measurement of the output qubit to the corresponding label in the data set.
Throughout this paper, we let this measurement be in the $Z$-basis. We now move on to describe our proposal.

\section{Methods}
\paragraph{Cost function embedding.\label{p_cf_embedding}}

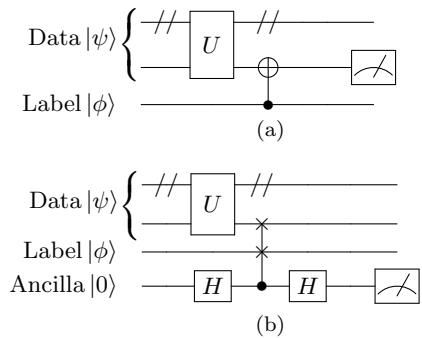
\begin{figure}
\centering
\subfloat[]{\label{fig:cost_cnot}
\Qcircuit @C=1em @R=0.7em {
\lstick{} & \qw // & \multigate{1}{U} & \qw   //     & \qw      & \qw & \qw   &\rstick{}     \\
\lstick{}& \qw  & \ghost{U}        & \targ     & \qw      & \qw  & \meter     \\
\lstick{\mathrm{Label} \ket{\phi}~} & \qw & \qw& \ctrl{-1} & \qw   & \qw    & \qw& \rstick{}
\inputgroupv{1}{2}{1em}{.7em}{\mathrm{Data} \ket{\psi} ~~~~~~}
}} 

\subfloat[]{\label{fig:cost_fredkin}
\Qcircuit @C=1em @R=0.7em {
\lstick{} & \qw // & \multigate{1}{U} & \qw   //     & \qw      & \qw   &\qw     \\
\lstick{}& \qw  & \ghost{U}        & \qswap     & \qw      & \qw   &\qw     \\
\lstick{\mathrm{Label} \ket{\phi}~} & \qw & \qw& \qswap \qwx& \qw      & \qw& \qw     \\
\lstick{\mathrm{Ancilla} \ket{0}~}& \qw & \gate{H}& \ctrl{-1}  & \gate{H} & \qw & \meter
\inputgroupv{1}{2}{1em}{.7em}{\mathrm{Data} \ket{\psi} ~~~~~~}
}}
\caption{Two ways to embeds cost functions into quantum circuits. The data qubits store the input data and label qubit stores the label (expected output). $U$ denote an arbitrary ansatz, while we consider the output from the ansatz is encoded in one qubit (here it is the qubit represented as the second line of the circuit). In Fig.\ref{fig:cost_cnot} we apply a CNOT gate on the output qubit, flip the state based on the label qubit. This will give us the CNOT cost function and encode the output into the Z-axis expectation value of the output qubit. In Fig.\ref{fig:cost_fredkin} we add an ancillary qubit and apply a Fredkin gate between the ansatz output and the label, controlled by the ancillary qubit. This will encode the overlap between ansatz output and the label into the ancillary qubit. 
}
\end{figure}

For the concrete implementation, we add another qubit $\ket{\phi}$, which holds the desired output for the supervised learning task, which we call the \textit{label qubit}.
For the binary classification problem, the label $\beta \in \{0,1\}$ can be encoded as states $\ket{0}$ or $\ket{1}$ of the label qubit.
A simple way to implement a cost function for binary classification is to use a CNOT gate to flip the output state based on the input training label, see Fig.\ref{fig:cost_cnot}.
Notably, under this operation, the output qubit remains $\ket{1}$ if the prediction and the expectation value are unequal, and will be $\ket{0}$ if they are the same.

Suppose the expectation value of the measurement $\braket{M}=\alpha$, and recall that the label is $\beta \in \{0,1\}$, then the CNOT cost function can be shown to be (see appendix \ref{app:CNOT_func})
\begin{equation}
    E_{cost(\psi)} = (1-2\beta)\alpha + \beta
\end{equation}

Note that this can be transformed into a well-studied loss function in ML, the so-called hinge loss~\cite{Duan2005WhichStudy}.

In general, we want to treat continuous as well as binary outcomes, i.e., $\mathcal{Y}=\mathbb{R}$. In the quantum setting, this means that the ancillary qubit would no longer be in one of the states $\{\ket{0},\ket{1}\}$, and the above CNOT trick can no longer be applied.
To handle this scenario, we make use of the well-known Swap Test \cite{Buhrman2001QuantumFingerprinting}. The Swap Test is a method to encode the overlap between two unknown states into an ancillary qubit. A swap test consists of a Hadamard gate followed by a controlled swap gate between two states and a final Hadamard gate.
If the two states that we aim to compare, let them be $\ket{\psi}$ and $\ket{\phi}$, are the same, i.e., $\braket{\psi | \phi}=1$, then the Swap Tests acts as the identity.
This means that the output state remains the same.
However, if the two states are different, i.e., $\braket{\psi | \phi} \neq 1$, then the difference will be reflected in the phase of the control qubit.
A swap test between the label state and the prediction state can hence be used to estimate the overlap.
We can then use the overlap as the output of our cost function, see Fig.~\ref{fig:cost_fredkin}.

Suppose the input state is $\ket{\psi}$, then the Z-axis expectation value of the measurement of the ancillary qubits can be written as \cite{Buhrman2001QuantumFingerprinting}

\begin{equation}
    E_{cost(\psi)} = \frac{1 - \bra{\phi}\rho_{output}\ket{\phi}}{2}
\end{equation}

The controlled swap gate is also referred to as the quantum Fredkin gate, and several physical realizations using photonics or superconducting systems have been realized.~\cite{Patel2016AGate, Ono2017ImplementationCircuits,Liu2018One-stepApplications}.
We hence have a single-qubit encoding of the cost function.

\paragraph{Data encoding.\label{p_data_encoding}}

Now let's consider the data encoding for the cost functions discussed above. In classical deep learning, the cost function is evaluated individually for each sampled data point from the training data set. Then the cost function value is averaged across all calculated values. This trivially costs $O(N)$ repetitions of the cost function evaluation for $N$ data points. Since we don't care about the individual value but only the averaged cost function value, we can investigate whether quantum parallelism can speed up its evaluation.

Consider a single data point. The goal of this procedure is to transfer each data point into a quantum state.
For fully digitized input data, i.e. if all the input values are binary (for example, 0 or 1), the encoding step can be done by mapping the classical ``1'' to the $\ket{1}$ state and classical ``0'' to the $\ket{0}$ state. 
Alternatively, if we would like to prepare non-digital input for  training, we first normalize it to the range $\left(0,\frac{\pi}{2}\right)$, and then encode the data in the amplitude of the input qubits.
This is done via the rotation angles, and we obtain for a single data point $j$ the angle $\gamma^j$ such that

\begin{equation}\label{eq:encode_1}
    \ket{\psi_i^j} = cos(\gamma_i^j)\ket{0} + sin(\gamma_i^j)\ket{1}
\end{equation}

Taking a data point in $N$ dimensions, the entire state is then given by the tensor product
\begin{equation}\label{eq:encode_2}
    \ket{\psi^j} = \bigotimes_{i=1}^N \ket{\psi_i^j}
\end{equation}
This method is often referred to as \textit{qubit encoding}~\cite{Stoudenmire2016SupervisedNetworks}.

Next, we need to consider how to represent our entire dataset and let the quantum circuit work out the averaged cost function naturally.

Suppose the circuit giving the cost function is represented by the unitary $C$, the measurement by the operator $O$, and the initial state for each data point be $\ket{\chi_i} = \ket{\psi_i}\otimes\ket{\phi_i}$. The cost function with respect to the $i$-th datapoint $x_i$ (encoded in $\psi_i$) can then be represented by 
 
 \begin{equation}
 [ E_{cost}]_{x_i} = \bra{\chi_i} C^\dag O C \ket{\chi_i} = \bra{\chi_i} D \ket{\chi_i},
 \end{equation}
where $D = C^\dag O C$. 
The average of the cost function is then given by 
\begin{equation}
 \overline{E_{cost}} = \frac{1}{N}\sum_{i=0}^N \bra{\chi_i} D \ket{\chi_i}
\end{equation}

where $N$ is the number of data points in the dataset.

Here we can use a mixed state, which is the classical average of all possible states of our datasets. The mixed state is simply the pure state constructed above with label qubits being traced away.
To see this, recall that the desired cost function value is the average of the cost function values across the entire data set.
If we hence input a mixed state,

\begin{equation} \label{eq:mix_state}
    \rho_{mix} = \frac{1}{n}\sum_n \ket{\phi_i}\ket{\psi_i}\bra{\psi_i}\bra{\phi_i},
\end{equation}
which is the average of all states in our data set, the outcome of the calculation is similarly the average over the data set.
This mixed state can be constructed by randomly selecting one of the samples in every single run of the algorithm, and then averaging the outcome over all runs. 
Since all the qubits are separable, we need single-qubit rotations for each qubit to prepare each state.
Notably, the accuracy of this will depend on the concrete distribution and the number of repetitions.

Rather than using the pure state described above, it is natural to think if we can prepare data in a pure superposition state. We find that it is possible, but that there is no speed up compared to the mixed state preparation method. Consider a uniform superposition state over the entire dataset. After evaluating the circuit, we will then obtain the expectation value as: 

\begin{equation}\label{eq:pure_state_derive_without_data}
    \begin{aligned}
 \overline{E_{cost}} =& \sqrt{\frac{1}{N}}\sum_{i=0}^N \bra{\chi_i} D \sqrt{\frac{1}{N}}\sum_{i=0}^N\ket{\chi_i} \\
    =& \frac{1}{N}\sum_{i=0}^N \bra{\chi_i} D \ket{\chi_i} + \frac{1}{N}\sum_{i=0}^N\sum_{j\neq i}^N \bra{\chi_i} D \ket{\chi_j}
    \end{aligned}
\end{equation}

The first term of Eq.~\ref{eq:pure_state_derive_without_data} would give us then exactly what we desire, if $i = j$. 
However, it also includes the terms for $i \neq j$.
Although the input states $\ket{\chi_i}$ are always orthogonal to each other for fully digitized input data, we cannot guarantee that $\bra{\chi_j}D\ket{\chi_i}=0$ holds.
To fix this problem, we can introduce an additional register $\ket{\epsilon_i}$ which we call the \textit{index qubits}.
The new initial state is then given by $\ket{\chi_i} = \ket{\psi_i}\otimes\ket{\phi_i}\otimes\ket{\epsilon_i}$.
We put the index of each data point into these qubits and don't apply any operation on them throughout the circuit. Since $\{\ket{\epsilon_i}\}$ is an orthogonal set, the entire state after running the circuit will also be orthogonal, and we will obtain $\bra{\chi_i} D \ket{\chi_j} = 0$ for any $i \neq j$. 

It is well known that such a superposition state can be prepared using variants of Grover's state preparation~\cite{Grover2000SynthesisComputation, Sanders2019Black-BoxArithmetic}. 
However, these approaches require relatively deep circuits, and are therefore not immediately applicable for NISQ devices.
Moreover, the index qubits we introduced here do not participate in any computation of the PQC. We can measure these index qubits at the very beginning, resulting in a state collapse into a specific data point. This superposition-state encoding is therefore equivalent to the mixed-state encoding we proposed above, which randomly selects a data-point at the beginning and averages the outcome in the end. This encoding will have the same outcome as long as the difference between the random number generator and the quantum randomness is negligible. 

\section{Results}

In this section we present numerical simulations of the proposed methods. 
We use the gradient-finding method using the Hadamard Test \cite{Dallaire-Demers2019Low-depthComputer, Romero2018StrategiesAnsatz} and the Adam optimizer \cite{Kingma2014Adam:Optimization}. First we give a brief review of the Hadamard Test gradient estimation, and then demonstrate how to use Hadamard Test gradient estimation together with the proposed method in this paper. Here we show two tasks, to train a classifier for a XOR gate and to train a classifier for a more realistic problem, the canonical Iris flower dataset.

\paragraph{Gradient estimation with the Hadamard test.\label{p_gradient_estimate}}

Gradient estimation with the Hadamard Test has previously been used to calculate the partial derivative of the eigenenergy of a molecule. \cite{Dallaire-Demers2019Low-depthComputer, Romero2018StrategiesAnsatz}
The Hamiltonian Pauli terms used to approximate molecular energies are typically multi-qubit terms. 
For this reason, they can require a separate control operation on a large number of qubits, which in NISQ devices has the potential to introduce a prohibitive level of noise \cite{Nielsen2000QuantumInformation}. Here, by encoding the cost function valuation into a single qubit, the control operation is reduced to a simple controlled-Z gate which is acceptable for NISQ applications.

\begin{figure*}[b]
    \centering
    \[\Qcircuit @C=1em @R=0.7em {
\lstick{} & \qw // &\multigate{1}{\mathcal{V}} &   \qw  //     &   \qw       &\multigate{1}{\mathcal{W}}    & \qw    & \qw        & \qw //        &   \qw     & \qw      & \qw & \qw\\
\lstick{}& \qw  &\ghost{\mathcal{V}}        &   \gate{e^{i\theta_j P}}       &  \gate{P}   &\ghost{\mathcal{W}}    &  \qw     & \targ       & \gate{Z}    & \qw      & \qw      & \qw& \qw\\
\lstick{\mathrm{Label} \ket{\phi}~}& \qw &\qw                        &   \qw       & \qw   \qwx  &   \qw                & \qw       & \ctrl{-1}  &   \qwx\qw   & \qw      & \qw      & \qw& \qw\\
\lstick{\mathrm{Ancilla} \ket{0}~}& \gate{H}& \qw              &   \qw       & \ctrl{-1}  &   \qw               & \qw        & \qw         & \ctrl{-2}  & \gate{H} &\qw & \gate{X\frac{\pi}{2}}      & \meter
\inputgroupv{1}{2}{1em}{.7em}{\mathrm{Data} \ket{\psi} ~~~~~~}\\
}\]
    \caption{Circuit with a CNOT cost function and single circuit gradient evaluation. The input data is encoded into data register $\psi$ represented as the first two lines, the expected output (label) is encoded in the label register $\ket{\phi}$. When we optimize the parameter $\theta_j$ we decompose the ansatz $U$ into $U = V(e^{i\theta_j P_j}\otimes \mathbb{I}_k^{(n-1)})W$ where $\mathbb{I}_k^{(n-1)}$ is the identity operation on all the other qubits except the $k$-th qubit (to which we applied the single qubit rotation), and $e^{i\theta_j P}$ is the single qubit rotation gate parametrized by $\theta_j$ where $P$ is a Pauli operator. We first insert a controlled-P gate right after the single qubit rotation, then add a controlled-Z gate to after the output of the cost function. Both these gates are controlled by an ancillary qubit prepared in state $\frac{1}{\sqrt{2}}(\ket{0}+\ket{1})$. In the end we rotate the ancillary qubit back to $\ket{0}\ket{1}$ basis, then do a half $\pi$ rotation on X axes. The partial derivative of the cost function with respect to $\theta_j$ is stored in the ancillary qubit. \label{fig:gradient_probe}}
\end{figure*}
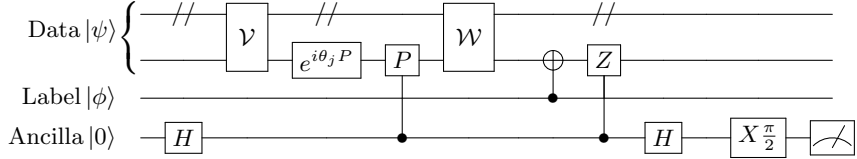

A circuit that implements this Hadamard Test gradient estimation together with cost function embedding is shown in Fig.~\ref{fig:gradient_probe}.
Using this method, it is possible to evaluate the partial derivative of the expectation value of the output qubit with respect to the rotation angle $\theta$. 
Here $\theta$ parametrizes the rotation generated by some Pauli operator $P$.
For example, for $P$ being the Pauli $X$ operator, we will have a single qubit rotation gate with angle $\theta$ about the $X$ axis. Typically there will be multiple single-qubit gates inside the ansatz. We can perform this method on each one of them to get the partial derivative of our cost function with respect to each rotation angle.
The whole process is given by the following:
We begin by preparing an ancillary qubit in the $\ket{0}$ state, and then apply a Hadamard gate to obtain the plus state, $\ket{+}=\frac{\ket{0} + \ket{1}}{\sqrt{2}}$.
Next, we apply a controlled-$P$ gate right after the single-qubit rotation gate $e^{i\theta P}$, where the control is on the ancillary qubit. 
Finally, because we encoded our cost function value in the Z-axis of the second qubit, we add a controlled-Z gate between the ancillary qubit and the second qubit, followed by another Hadamard gate on the ancillary qubit which rotates the state back to the $\{0,1\}$-basis.
The resulting ancillary qubit now contains the gradient encoded in its phase. 
Finally, we apply a $\frac{\pi}{2}$ rotation about the X-axis on the ancillary qubit so that the gradient is encoded in the Z-axis of the ancilla, and can be determined by measuring in the $\{\ket{0},\ket{1}\}$ basis.

We can therefore use the circuit in Fig.~\ref{fig:gradient_probe} to estimate the gradient by simply measuring the single ancillary qubit.

\paragraph{XOR experiment}

We can now proceed to test the above methods for training PQCs.

\begin{figure*}[b]
    \centering

    \[\Qcircuit @C=.5em @R=1em {
\lstick{} &\qw&\qw&\qw&\qw&\qw&\qw &\qw&\qw&\qw &\qw &\qw&\qw&\qw\\
\lstick{} &\qw&\qw&\qw&\qw&\qw&\qw &\qw&\qw&\qw &\qw &\qw&\qw&\qw\\
\lstick{} &\gate{R_x}&\gate{R_z}&\gate{R_x}&\ctrl{1}&\qw&\qw&\qw&\qw&\qw&\qw&\qw&\qw&\qw\\
\lstick{} &\gate{R_x}&\gate{R_z}&\gate{R_x}&\targ&\gate{R_x(\theta)}&\targ&\gate{R_z}&\gate{R_x}&\targ&\gate{Z}&\qw&\qw&\qw\\
\lstick{\mathrm{Label} \ket{\phi}~} &\qw&\qw&\qw&\qw&\qw&\qw \qwx&\qw&\qw&\ctrl{-1}&\qw \qwx&\qw&\qw&\qw\\
\lstick{\mathrm{Ancilla} \ket{0}~} &\gate{H}&\qw&\qw&\qw&\qw&\ctrl{-2}&\qw&\qw&\qw&\ctrl{-2}&\gate{H}&\gate{X\frac{\pi}{2}}&\meter
\inputgroupv{3}{4}{1em}{1em}{\mathrm{Data} \ket{\psi} ~~~~~~}
\inputgroupv{1}{2}{1em}{1em}{\mathrm{Index} \ket{\epsilon} ~~~~~~}
}\]
    \caption{Evaluation of the partial derivative of theta for training the XOR circuit with CNOT cost function. This circuit is used to find the partial derivative of the CNOT cost function with respect to the rotation angle $\theta$. Here we choose an ansatz with arbitrary single qubit rotation applied on each input qubit first, then use a CNOT gate to exchange the information between two qubits, and again apply an arbitrary single qubit rotation on the output qubit. We use three parametrized single qubit rotations, \textit{Rx}, \textit{Rz}, \textit{Rx}, to construct an arbitrary single qubit rotation. The input state is encoded with equation \ref{eq:xor_pure_state}. The index qubit is initialized as $\ket{00}$ and the first two CNOTs acting on index is used to copy the data qubits into the index qubits, we can achieve the copy because we know the data qubits are all in $\ket{0} \ket{1}$ basis. \label{fig:CNOT_XOR_pure_state}}
    
\end{figure*}
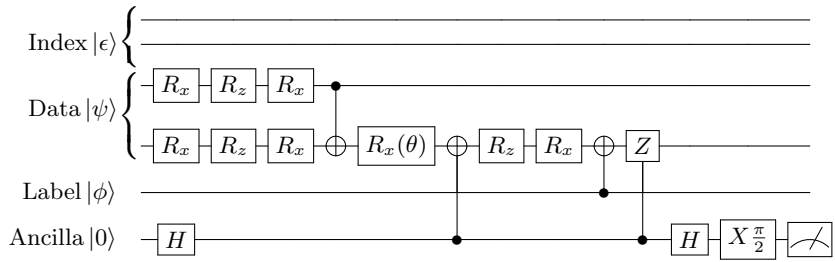

The first example task consists of training the circuit to perform the classical exclusive or (XOR) operation. 
Here we require the circuit to implement the truth table of XOR, i.e., yield true if and only if the input bits are different, and false otherwise. 
We use $\ket{0}$ to denote the FALSE value, and $\ket{1}$ to denote TRUE. Here we demonstrate the method encoding data in superposition. The circuit ansatz is shown in Fig.\ref{fig:CNOT_XOR_pure_state}.

The input bits are $\ket{{\color{blue}00}}$, $\ket{{\color{blue}01}}$, $\ket{{\color{blue}10}}$, $\ket{{\color{blue}11}}$, the corresponding labels ,i.e., evaluations of XOR on the respective input, are $\ket{{\color{teal}0}}$, $\ket{{\color{teal}1}}$, $\ket{{\color{teal}1}}$, $\ket{{\color{teal}0}}$, and the indices are given by $\ket{{\color{red}00}}$, $\ket{{\color{red}01}}$, $\ket{{\color{red}10}}$, $\ket{{\color{red}11}}$. 
So the final input state is given by 

\begin{equation}\label{eq:xor_pure_state}
    \ket{\chi} = \frac{1}{2}(
    \ket{{\color{red}00}{\color{blue}00}{\color{teal}0}0}+
    \ket{{\color{red}01}{\color{blue}01}{\color{teal}1}0}+
    \ket{{\color{red}10}{\color{blue}10}{\color{teal}1}0}+
    \ket{{\color{red}11}{\color{blue}11}{\color{teal}0}0})
\end{equation}

\begin{figure*}[t]
    \centering
    \includegraphics[width=.9\textwidth]{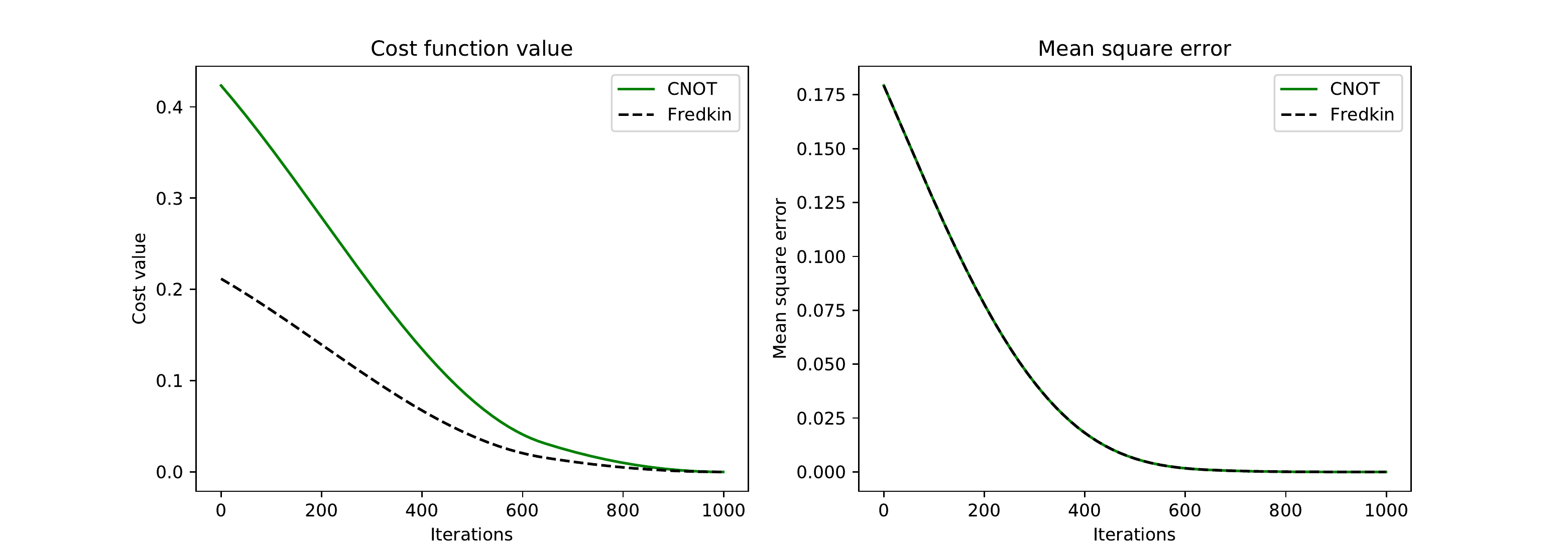}
    \caption{Numerical simulation result for circuit showed in figure \ref{fig:CNOT_XOR_pure_state}. Here we tested the cost function value and the mean square error for the same training dataset prepared as equation \ref{eq:xor_pure_state}. During the training process, the circuits for evaluating partial derivative of each the parameters $\theta_i$ are generated. For each iteration, all these circuits are evaluated and gives the derivative of each parameters, which is the gradient of the cost function. Then the gradient value is passed to the Adam optimizer. Here the learning rate of the optimizer is $10^{-3}$. At the end of each iteration, the parameters are updated by the Adam optimizer. The result here shows both CNOT and Fredkin cost function can be used for training, and they give similar convergence speed.\label{fig:xor_analysis}}
\end{figure*}

After the gradients are determined by evaluating the circuit they are used to update the parameters using the Adam optimizer \cite{Kingma2014Adam:Optimization}. 

The simulation result is shown in Fig.\ref{fig:xor_analysis}. 
During the experiment, we were able to see that circuits with CNOT or Fredkin cost functions can successfully give us the correct gradient direction. 
We observed that training converges for both fully digitized data encoding or the mixed state encoding, and the convergence rates are the same.

\paragraph{Iris experiment}

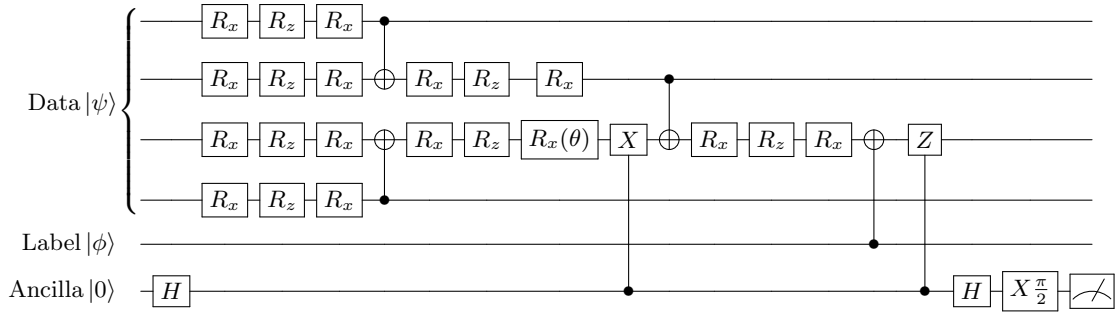
\begin{figure*}[t]
\centering
\[\Qcircuit @C=.5em @R=1em {
\lstick{} & \qw     &\gate{R_x}&\gate{R_z}&\gate{R_x} & \ctrl{1}  & \qw      & \qw      & \qw      & \qw      & \qw      & \qw      & \qw      &\qw        &\qw        &\qw       &\qw      &\qw    &\qw       &\qw        &\rstick{}\\
\lstick{} & \qw     &\gate{R_x}&\gate{R_z}&\gate{R_x} & \targ     &\gate{R_x}&\gate{R_z}&\gate{R_x}& \qw      &\ctrl{1}  & \qw      & \qw      &\qw        &\qw        &\qw       &\qw      &\qw     &\qw       &\qw       &\rstick{}\\
\lstick{} & \qw     &\gate{R_x}&\gate{R_z}&\gate{R_x} & \targ     &\gate{R_x}&\gate{R_z}&\gate{R_x(\theta)}& \gate{X} &\targ     &\gate{R_x}&\gate{R_z}&\gate{R_x} & \targ     &\qw       &\gate{Z} &\qw     &\qw       &\qw       &\rstick{}\\
\lstick{} & \qw     &\gate{R_x}&\gate{R_z}&\gate{R_x} & \ctrl{-1} & \qw      & \qw      & \qw      & \qw\qwx  & \qw      & \qw      & \qw      &\qw        &\qw        &\qw       &\qw      &\qw     &\qw       &\qw       &\rstick{}\\
\lstick{\mathrm{Label} \ket{\phi}~} & \qw     &  \qw     &  \qw     &  \qw      &  \qw      &  \qw     & \qw      & \qw      & \qw      & \qw      & \qw      & \qw      &\qw       &\ctrl{-2}  &\qw       &\qw      &\qw      &\qw       &\qw      &\rstick{}   \\
\lstick{\mathrm{Ancilla} \ket{0}~} & \gate{H}&  \qw     &  \qw     &  \qw      &  \qw      &  \qw     & \qw      & \qw      & \ctrl{-2}& \qw      & \qw      & \qw      &\qw        &\qw       &\qw       &\ctrl{-3} &\gate{H}   &\gate{X\frac{\pi}{2}} & \meter      
\inputgroupv{1}{4}{1em}{3.4em}{\mathrm{Data} \ket{\psi} ~~~~~~}
}\]
\caption{Training circuit for quantum hierarchical classifiers \cite{Grant2018HierarchicalClassifiers}. The input state is encoded with equation \ref{eq:mix_state}. Here we demonstrate that the circuit gives the partial derivative of the cost function against the parameter $\theta$ which parametrize one of the single qubit rotation gate. A controlled X gate (CNOT) is applied right after the parametrized gate.\label{fig:qcircuit_iris}}
\end{figure*}

Next we investigate a more realistic problem. As a second example we implement a classifier for Iris dataset~\cite{FISHER1936THEPROBLEMS}. 
The Iris dataset contains 150 labelled examples in total, with three different types of Iris flowers. Each example is described by four features. 
We prepare a mixed state with the method from Eq. \ref{eq:mix_state}, and the training circuit is shown in Fig.\ref{fig:qcircuit_iris}.

\begin{figure*}[t]
    \centering
    \includegraphics[width=.9\textwidth]{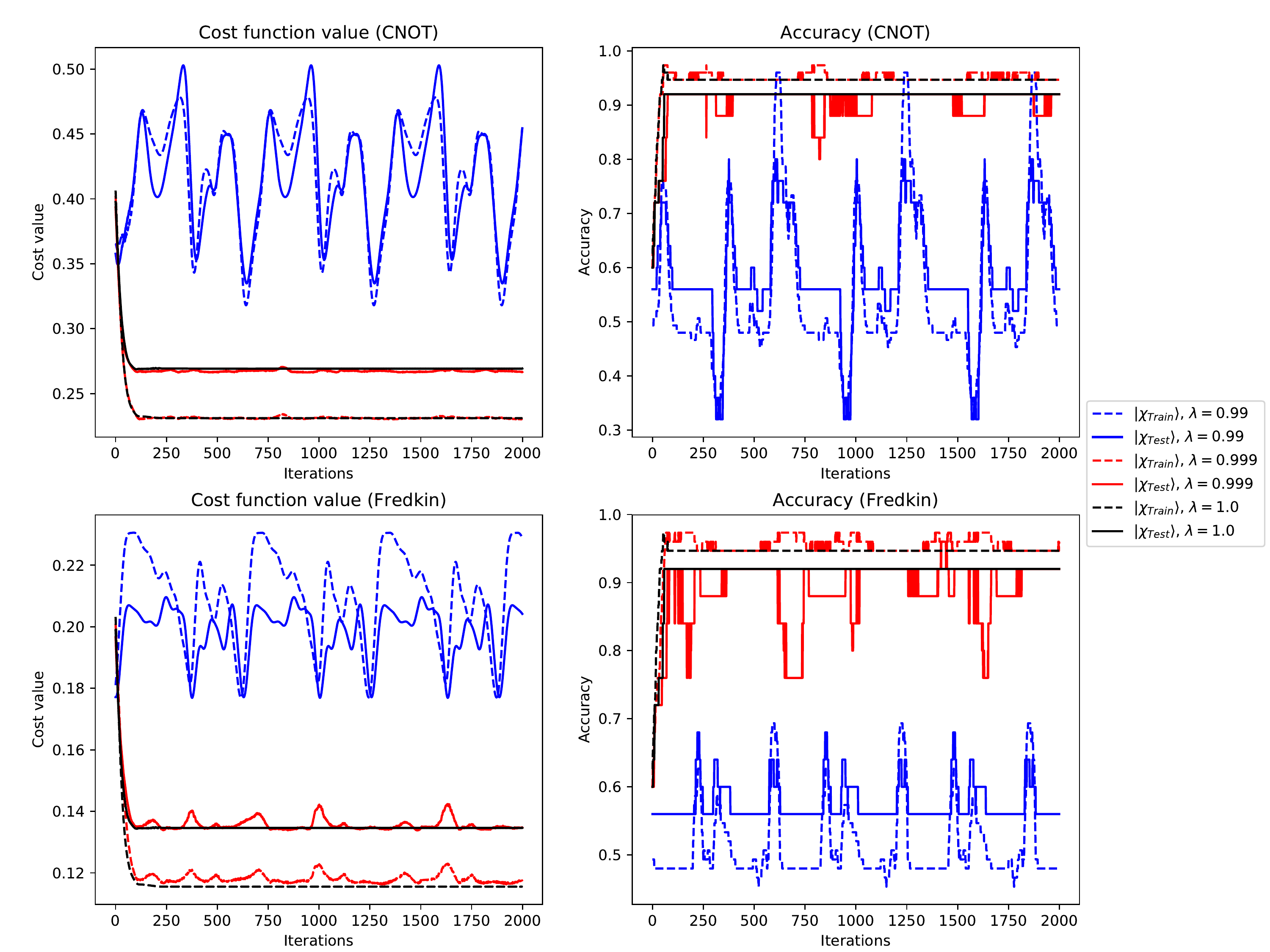}
    \caption{Simulation of training the hierarchical classifier for Iris dataset. Here we choose two flowers (label 1 and 2) from the Iris dataset. The training process is the same as Fig.\ref{fig:xor_analysis}. Here the learning rate is $10^{-2}$. We consider the prediction of the flower label is 2 if the probability to measure and get state $\ket{1}$ is greater than 0.5, otherwise the prediction will be considered to be label 1. The accuracy is the ratio of the correct prediction among the entire training dataset $\ket{\chi_{Train}}$ (dashed lines) or the test dataset $\ket{\chi_{Test}}$ (solid lines). Then we add the depolarization channel $\Delta_\lambda(\rho)=\lambda \rho + (1-\lambda)/2^n \mathbb{I}$ where $n$ is the number of qubits to the circuit and compare the simulation between results with or without noise.\label{fig:iris_analysis} }
\end{figure*}

Several interesting results were found during the training process, see Fig.\ref{fig:iris_analysis}. First, we report that by using Adam optimizer, the PQCs can converge. The CNOT cost function and the Fredkin cost function were able to achieve similar training performance and convergence rates. 

In order to investigate the robustness of the circuit, we applied the depolarization channel to the system. The channel is described as $\Delta_\lambda(\rho)=\lambda \rho + (1-\lambda)/2^n \mathbb{I}$ where $n$ is the number of qubits. 
We observe that the algorithm still converges to a region close to the optimal point when $\lambda =0.999$ is applied. By recording the value of parameters and the value of cost functions, we can choose a threshold for the error to be used as a stopping criterion for training. When $\lambda=0.99$, the cost function no longer converges.

 We can see for the circuit in Fig.\ref{fig:qcircuit_iris} that we can tolerate depolarization noise when $\lambda = 0.999$, which is quite close to gate fidelities of state-of-art NISQ devices \cite{Klimov2018FluctuationsQubits, Linke2017ExperimentalArchitectures}. This method could also be combined with error mitigation \cite{Endo2018PracticalApplications}, which can suppress some errors and make this algorithm even more robust to the noise on a NISQ machine.

\section{Conclusion}

We now summarize the advantages of our proposed method. Firstly, the encoding of the value of the cost function into a measurement expectation value enables the ability to do further quantum information processing. For example, the direct usage of advanced optimization methods such as the Hadamard Test, \textit{shift-rule}, and \textit{rotosolve}. Since the cost function value is encoded in the state of a single qubit, the implementation of a full optimization algorithm making use of this method can be straightforward. 

Secondly, the proposed data encoding method allows for efficiently estimating expectation values across the training dataset. It is classically inefficient to calculate the cost function for each example and then average them across the entire data set. Instead, the proposed method indicates that the expectation value can be obtained via random sampling from the training dataset.

We show that simply encoding the dataset into a superposition state will not give us the average of the cost function values. To fix this issue, we introduced the index qubit. However, we showed that index qubit would make it equivalent to doing a mixed state preparation and that such superposition state preparation has no speed-up compare to mixed state preparation.

In conclusion, we have proposed a method to encode cost functions into quantum circuits and a corresponding method for preparing the input data. This method therefore enables quantum information processing on the cost function value. The averaged cost value can be calculated across the entire dataset with both mixed state data preparation and superposition data preparation. We demonstrated gradient evaluation of the cost function with the Hadamard Test and investigated its performance under a depolarization noise channel.

\vspace{1em}

\section*{Acknowledgements}

E.G.~is supported by the UK Engineering and Physical Sciences Research Council (EPSRC) [EP/P510270/1]. L.W.~acknowledges the support through the Google PhD Fellowship in Quantum Computing. B.V.~acknowledges support from an EU Marie Skłodowska-Curie fellowship. P.L.~acknowledges support from the EPSRC [EP/M013243/1] and Oxford Quantum Circuits Limited. We thank M.~Benedetti, X.~Yuan, P.~Spring and T.~Tsunoda and for insightful discussions. We acknowledge the use of the University of Oxford Advanced Research Computing facility.

\bibliography{main}

\appendix
\begin{widetext}

\section{CNOT Cost function derivation}\label{app:CNOT_func}

Suppose the expectation value of the measurement $\braket{M}=\alpha$, the state of the output qubit is given by 

\begin{equation}
\rho_0 = \begin{pmatrix} 
1-\alpha & -\epsilon^* \\
\epsilon & \alpha 
\end{pmatrix}
\end{equation}

For each data point, the label qubit can be in state either $\ket{0}$ or $\ket{1}$, which we introduce $\beta = 0$ when state $\ket{0}$ and $\beta = 1$ when state $\ket{1}$. The state of the label qubit is given by

\begin{equation}
\rho_{\phi} = \begin{pmatrix} 
1-\beta & 0 \\
0 & \beta 
\end{pmatrix}
\end{equation}

And the state of the system of labeling qubit and output would be 

\begin{equation}
\rho_1 = \rho_0 \otimes \rho_{\phi} = \begin{pmatrix} 
 (1-\alpha ) (1-\beta ) & 0 & -(1-\beta ) \epsilon ^* & 0 \\
 0 & (1-\alpha ) \beta  & 0 & -\beta  \epsilon ^* \\
 (1-\beta ) \epsilon  & 0 & \alpha  (1-\beta ) & 0 \\
 0 & \beta  \epsilon  & 0 & \alpha  \beta  \\
\end{pmatrix} 
\end{equation}

After apply the CNOT gate, the state would be

\begin{equation}
\rho_2 = U^\dag_{CNOT} \rho_1 U_{CNOT} = \begin{pmatrix}
 (1-\alpha ) (1-\beta )  & 0                   & -(1-\beta ) \epsilon ^* & 0 \\
 0                       & \alpha  \beta       & 0                       & \beta  \epsilon  \\
 (1-\beta ) \epsilon     & 0                   & \alpha  (1-\beta )      & 0 \\
 0                       & -\beta  \epsilon ^* & 0                       & (1-\alpha ) \beta  \\
\end{pmatrix}
\end{equation}

Trace away the labeling qubit, the state of the output qubit would be

\begin{equation}
\rho_{output} = Tr_{\phi}(\rho_2) = 
\begin{pmatrix} 
 (1-\alpha ) (1-\beta ) + \alpha  \beta  & -(1-\beta ) \epsilon ^* +  \beta  \epsilon \\
(1-\beta ) \epsilon -\beta  \epsilon ^* & \alpha  (1-\beta ) + (1-\alpha ) \beta
\end{pmatrix}
\end{equation}

The expectation value of the output qubit would be

\begin{equation}
\begin{aligned}
    E_{cost(\psi)} &= P_{\ket{1}} \\
    &= \alpha  (1-\beta ) + (1-\alpha ) \beta \\
    &= \alpha + \beta -2\alpha\beta\\
    &= (1-2\beta)\alpha + \beta
\end{aligned}
\end{equation}

\end{widetext}

\end{document}